\documentclass[12pt]{iopart}
\usepackage[dvips]{graphicx}
\usepackage{amssymb}

\begin{document}

\title{On the physics behind the form factor ratio
$\mu_p \,G_E^p (Q^2) / G_M^p (Q^2)$}

\author{
M.~Wakamatsu and Y.~Nakakoji
}
\address{
Department of Physics, Faculty of Science, \\
Osaka University, \\
Toyonaka, Osaka 560-0043, JAPAN
}
\ead{wakamatu@phys.sci.osaka-u.ac.jp ; nakakoji@kern.phys.sci.osaka-u.ac.jp}

\begin{abstract}

We point out that there exist two natural definitions of the
nucleon magnetization densities : the density $\rho_M^K (r)$
introduced in Kelly's phenomenological analysis and
theoretically more standard one $\rho_M (r)$.
We derive an explicit analytical relation between them, although
Kelly's density is more useful to disentangle
the physical origin of the different $Q^2$ dependence of the Sachs
electric and magnetic form factors of the nucleon.
We evaluate both of $\rho_M (r)$ and $\rho_M^K (r)$ as well as the
charge density $\rho_{ch}(r)$ of the proton within the framework
of the chiral quark soliton model, to find a noticeable qualitative
difference between $\rho_{ch}(r)$ and $\rho_M^K (r)$, which is
just consistent with Kelly's result obtained from the empirical
information on the Sachs electric and magnetic form factors
of the proton.

\end{abstract}

\maketitle

The electromagnetic form factors are one of the most fundamental
observables, which characterize the underlying composite structure
of the nucleon \cite{HW04}\nocite{Gao05}\nocite{ARZ07}-\cite{PPV07}.
As is widely known, the expectation from perturbative
QCD (pQCD) is that the $Q^2$ dependence of the Sachs electric, $G_E (Q^2)$,
and magnetic, $G_M (Q^2)$, form factors should be the same at large
$Q^2$ \cite{BF73},\cite{MMT73}, and early experimental data obtained
by the standard Rosenbluth technique appeared to be qualitatively
consistent with this expectation \cite{Walker94}.
However, the recent experiments at Jefferson Lab,
utilizing the polarization transfer technique found the surprising
fact that $G_E (Q^2)$ decreases more rapidly than $G_M (Q^2)$
at large $Q^2$ \cite{JLab01}\nocite{JLab02}-\cite{JLab05}.
A number of theoretical analyses carried out since then,
indicate that the discrepancy between the two different techniques for
extracting form factor ratio is most likely to be resolved if
the two-photon-exchange contributions in elastic $ep$
scatterings are taken into account \cite{GV03}\nocite{BMT03}\nocite{CABCV04}
\nocite{KMBT05}\nocite{BM06}\nocite{CV07}\nocite{KB07}-\cite{AMT07},
thereby providing a strong support to the discovery by the JLab
measurements. (For more detail, see, for example, the recent global
analysis of the nucleon form factors by Arrington, Melnitchouk and Tjon,
and references therein \cite{AMT07}.)

An interesting theoretical challenge is therefore how we can understand
the physics behind this remarkable
observation \cite{Kelly02}\nocite{LYT00}\nocite{Miller02}\nocite{BGKPRW02}
\nocite{FW03}\nocite{FGLP06}\nocite{BHM07}\nocite{PB07}
\nocite{GRP08}\nocite{Holz96}\nocite{Holz02}-\cite{Holz05}.
Natural objects of study here would be charge and magnetization densities,
which are defined as Fourier transforms of the Sachs electric
and magnetic form factors in the Breit frame. Here is a subtlety, however.
The problem is that the Breit frame varies with $Q^2$
so that the electromagnetic densities so defined are frame-dependent
quantities.
To obtain electromagnetic densities in the nucleon rest frame, which
have intrinsic physical meaning, Kelly proposed to introduce
what-he-calls the {\it intrinsic form factors} defined as the
Fourier transforms of the charge and magnetization densities at
the nucleon rest frame \cite{Kelly02}.
It is assumed that these intrinsic form factors are directly related to
the measured Sachs electric and magnetic form factors on account
of some relativistic effects.
(The most important relativistic effect is Lorentz contraction of
spatial distributions in the Breit frame.)
This allows him to extract the intrinsic charge and magnetization
densities, $\rho_{ch}(r)$ and $\rho_M^K (r)$ in the nucleon rest frame
from the available empirical information on the Sachs electric and
magnetic form factors. The result of his analysis illustrated in Fig.7
of \cite{Kelly02} clearly shows that the peak of the $r^2$-weighted magnetic
density, i.e. $r^2 \,\rho_M^K (r)$ is located at smaller $r$ than the
corresponding peak of the charge density $r^2 \,\rho_{ch}(r)$, which is
thought to explain faster falloff of the electric form factor
as compared with the magnetic one.

Now, the purpose of our present study is to give further theoretical
support to Kelly's phenomenological findings.
Our strategy to accomplish it is as follows.
First, we predict the intrinsic charge and magnetization densities
defined by Kelly, within the framework of the chiral quark soliton model
(CQSM) \cite{DPP88},\cite{WY91}. (For early reviews of the CQSM, see
\cite{CQSMreview1}\nocite{CQSMreview2}-\cite{CQSMreview3}.)
We immediately notice that the magnetization density that naturally
appear in the standard theoretical framework is different from the
corresponding magnetization density introduced by Kelly.
However, the point of our analysis is that we can readily derive
an analytical relation between them.
This then enables a direct calculation of the intrinsic charge
and magnetization densities {\it a la} Kelly within a single
theoretical framework, {\it without worrying about the Lorentz Boost
effects}. The comparison of these two intrinsic form factors is then expected
to provide us with a valuable hint for clarifying the physical origin of the
different $Q^2$ behavior of the observed charge and magnetic form factors.
Furthermore, to confirm that the predicted delicate difference between the
intrinsic charge and magnetization densities is in fact an
essential ingredient to explain the observed difference
between the Sachs electric and magnetic form factors,
we also evaluate the latters explicitly by taking account of the
Lorentz boost effects with the simplest prescription proposed
in the past studies.

There have already been several investigations of the nucleon
electromagnetic form factors within the framework of the
CQSM \cite{Waka91}\nocite{Waka92}\nocite{CGG95}-\cite{KWG97}.
In all these studies,
however, the treatment of the nucleon center-of-mass motion is
essentially nonrelativistic, which means that the reliability of
the theoretical predictions is limited to the low-momentum-transfer
domain $Q^2 \ll M_N^2$ with $M_N$ being the nucleon mass.
Although a complete relativistic treatment of a bound state
is an extremely difficult problem especially for a field
theoretical model like the CQSM, there is an approximate way, adopted
by Kelly's phenomenological analysis, to implement the relativistic
recoil corrections, or equivalently the
effects of Lorentz boost from the rest frame to the Breit frame.
(By the term field theoretical model, here we mean a model of the
nucleon, which contains not only the lowest $q^3$ (3-quark) Fock space
but also higher $q^3 \,(\bar{q} q)^n$ Fock spaces.)
In this prescription, one first introduces the intrinsic charge,
$\tilde{\rho}_{ch} (k)$, and magnetic, $\tilde{\rho}_m (k)$, form
factors through the relations \cite{Kelly02} : 
\begin{eqnarray}
 \tilde{\rho}_{ch} (k) &=& G_E (Q^2) \,
 \left( 1 + \tau \right)^{\lambda_E} , \label{eq:boost1} \\
 \mu_p \,\tilde{\rho}_m (k) &=& G_M (Q^2) \,
 \left( 1 + \tau \right)^{\lambda_M} , \label{eq:boost2}
\end{eqnarray}
with $G_E (Q^2)$ and $G_M (Q^2)$ corresponding to the observed electric
and magnetic form factors. In the above equations,
$k$ is the intrinsic spatial frequency corresponding to the
momentum transfer $Q^2$ : 
\begin{equation}
 k^2 \ = \ Q^2 \,/ \,(1 + \tau) , \label{eq:boost3}
\end{equation}
where $\tau = Q^2 \,/ \,(4 \,M_B^2)$ with $M_B$ being the boost mass,
which ideally should coincide with the physical nucleon mass.
The parameters $\lambda_E$ and
$\lambda_M$ are integers, whose values are model dependent.
In Kelly's phenomenological analysis, he used the choice
$\lambda_E = \lambda_M = 2$ \cite{Kelly02},
which was first suggested by Mitra and Kumari in the cluster
model \cite{MK77}.
(For other choices, see \cite{Kelly02},\cite{Holz96}
\nocite{Holz02}-\cite{Holz05},\cite{MK77}\nocite{LP70}-\cite{Ji91},
for instance.)
Kelly then defines the intrinsic charge and magnetization
densities as Fourier transforms of the above intrinsic form
factors \cite{Kelly02} : 
\begin{eqnarray}
 \rho_{ch} (r) &=& \frac{2}{\pi} \,\int_0^\infty \,
 d k \,k^2 \,j_0 (kr) \,\tilde{\rho}_{ch} (k), \label{eq:chkdens} \\
 \rho_m^K (r) &=& \frac{2}{\pi} \,\int_0^\infty \,
 d k \,k^2 \,j_0 (kr) \,\tilde{\rho}_m (k).
 \label{eq:mgkdens}
\end{eqnarray}
They are the quantities to be identified with the static densities
in the nucleon rest frame, so that they can be predicted, for instance,
by the CQSM, or by any other models, without any difficulty.
(This statement is true for the charge density, but
the magnetization density defined above is different from more standard
one, which appears naturally in the theoretical formula of the
Sachs magnetic form factor. See the discussion below.)
Since the theoretical expressions for the charge and magnetic form
factors within the framework of the CQSM were already given in
several previous papers \cite{CGGP95}\nocite{KWG97}-\cite{WN06},
we recall here only their general
theoretical structures, i.e. the dependence on the collective angular
velocity $\Omega$ of the rotating soliton, which scales as
$1 / N_c$ \cite{DPP88}.
By definition, the intrinsic charge form factor $\tilde{\rho}_{ch}(k)$
is the Fourier transform of the intrinsic charge density
$\rho_{ch}(r)$ :

\begin{equation}
 \tilde{\rho}_{ch} (k) \ = \ \int_0^\infty \,dr \,r^2 \,
 j_0 (kr) \,\rho_{ch} (r),
 \label{eq:chdens}
\end{equation}
Note that the intrinsic charge density $\rho_{ch}(r)$ is an easily
tractable theoretical object, since it can be calculated in the
nucleon rest frame, without worrying about the effects of
Lorentz boost. Within the CQSM, the intrinsic charge density of the
proton $\rho_{ch}(r)$ is obtained as the sum of the isoscalar and the
isovector parts as
\begin{equation}
 \rho_{ch}(r) \ = \ \rho_{ch}^{(I=0)}(r) \ + \ \rho_{ch}^{(I=1)}(r).
\end{equation}
The isoscalar part receives the zeroth order contribution in the collective
angular velocity $\Omega$ of the soliton, while the isovector part survives
only at the 1st order in $\Omega$ : 
\begin{equation}
 \rho_{ch}^{(I=0)}(r) \ \sim \ O(\Omega^0), \ \ \ \ \ 
 \rho_{ch}^{(I=1)}(r) \ \sim \ O(\Omega^1).
\end{equation}
On the other hand, the intrinsic magnetic form factor is given in
the form : 
\begin{equation}
 \mu_p \,\tilde{\rho}_m (k) \ = \ \mu_p \,\int_0^\infty \,dr \,r^2 \,\,
 \frac{3 \,j_1 (kr)}{kr} \,\,\rho_m (r),
 \label{eq:mgdens}
\end{equation}
with
\begin{equation}
 \rho_m (r) \ = \ \rho_m^{(I=0)} (r) \ + \ \rho_m^{(I=1)} (r).
\end{equation}
Here, the isoscalar part survives only at the 1st order in
$\Omega$, while the isovector part consists of the leading-order
term and the 1st order rotational correction : 
\begin{equation}
 \rho_m^{(I=0)}(r) \ \sim \ O(\Omega^1), \ \ \ \ \ 
 \rho_m^{(I=1)}(r) \ \sim \ O(\Omega^0) \ + \ O(\Omega^1) .
\end{equation}
(It is known that the existence of the 1st order rotational correction
in the CQSM, which is absent in the Skyrme type
effective meson theories, is essential for reproducing the correct
magnitude of the isovector magnetic moment of the nucleon as well as
that of the isovector axial
charge \cite{WW93}\nocite{CBGPPWW04}\nocite{Waka96}-\cite{WN07}.)
Here, for the sake of comparison to be made later,
not only the intrinsic charge density in Eq.(\ref{eq:chdens}) but also
the intrinsic magnetization density in Eq.(\ref{eq:mgdens}) is defined 
so that they satisfy the normalization conditions : 
\begin{equation}
 1 \ = \ \int_0^\infty \,dr \,r^2 \,\rho_{ch} (r) \ = \ 
 \int_0^\infty \,dr \,r^2 \,\rho_m (r).
\end{equation}
An important notice here is that the intrinsic magnetization density
appearing in Eq.(\ref{eq:mgdens}) is different from the corresponding
Kelly's magnetization density $\rho_m^K (r)$ appearing in
Eq.(\ref{eq:mgkdens}).
The magnetization density $\rho_m (r)$
appears naturally in a theoretical formula for the magnetic
form factor. On the other hand, as is obvious from
Eqs.(\ref{eq:chkdens}) and (\ref{eq:mgkdens}),
Kelly's magnetization density is defined completely in parallel with
the charge density, as a Fourier-Bessel transform of the intrinsic magnetic
form factor, so that it is more useful to unravel the coordinate space origin
of the different behaviors of the Sachs electric and magnetic form
factors. Comparing their definitions, however, it is an easy exercise to
derive the relation between these two magnetization densities.
Using the identity of the spherical Bessel function
\begin{equation}
 \int_0^\infty \,j_l (k r_1) \,j_{l+1} (k r_2) \,\,k \,dk \ = \ 
 \frac{\pi}{2} \,\,\frac{r_1^l}{r_2^{l+2}} \,\,\theta (r_2 - r_1) ,
\end{equation}
we readily find the desired relation
\begin{equation}
 \rho_m^K (r) \ = \ 3 \,\int_r^\infty \,d r^\prime \,\,
 \frac{\rho_m (r^\prime)}{r^\prime} ,
 \label{eq:mgrelation}
\end{equation}
or, inversely
\begin{equation}
 \rho_m (r) \ = \ - \,\frac{1}{3} \,r \,\frac{d}{d r} \,\rho_m^K (r).
\end{equation}
Note that $\mu_p \,\rho_m (r)$ and $\mu_p \,\rho_m^K (r)$ would have
totally different radial dependence, but they have a common
normalization,
\begin{equation}
 \mu_p \ = \ \int_0^\infty \,dr \,r^2 \,\mu_p \,\,\rho_m (r) \ = \ 
 \int_0^\infty \,dr \,r^2 \,\mu_p \,\,\rho_m^K (r),
\end{equation}
so that both are qualified to be called the magnetization density of
the proton.

Now, we are ready to show what the predictions of the CQSM are like.
For the regularization scheme, we use here the Pauli-Villars one with
double subtraction proposed in \cite{KWW99}.
Setting the pion mass to the physical value, i.e. $138 \,\mbox{MeV}$,
the model contains only one
parameter, i.e. the dynamical quark mass $M$, playing the role of
quark-pion coupling strength. Favorable physical
predictions of the model were known to be obtained with use of
the dynamical quark mass $M \simeq (375 - 400) \,\mbox{MeV}$ \cite{WK99}.
Since the purpose of our present study is not to precisely reproduce
the observed electromagnetic form factors of the nucleon but to
understand the origin of the remarkable qualitative difference between
the observed Sachs electric and magnetic form factors, we simply use the
value $M = 375 \,\mbox{MeV}$ in the following.
(Our strategy should, for instance, be contrasted with Holzwarth's
analysis \cite{Holz96}\nocite{Holz02}-\cite{Holz05}, which
introduces many adjustable parameters to precisely reproduce the observed
electromagnetic form factors of the nucleon, although based on a
similar soliton model, i.e. the generalized Skyrme model with
vector mesons.)

\begin{figure}[htb]
\begin{center}
  \includegraphics[height=.36\textheight]{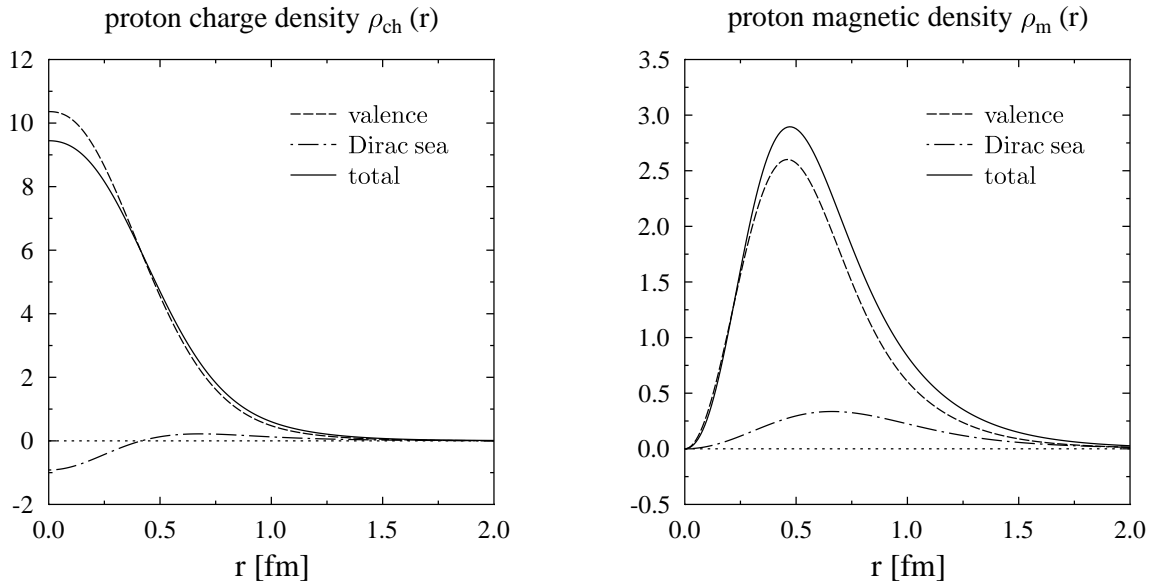}
  \caption{\baselineskip16pt The CQSM predictions for the intrinsic
  charge (left panel) and magnetization (right panel) densities defined
  in Eqs.(\ref{eq:chdens}) and (\ref{eq:mgdens}).}%
\label{Fig1}
\end{center}
\end{figure}

Almost parameter free predictions of the CQSM for the intrinsic
charge and magnetization densities are shown in Fig.\ref{Fig1}.
The magnetization density shown here is the standard one appearing
in Eq.(\ref{eq:mgdens}).
In both panels, the dashed and dash-dotted curves stand for
the contribution of the $N_c \,(=3)$ valence quarks and that of the
deformed Dirac-sea quarks, while their sums are shown by the solid
curves. A stronger effect of the Dirac-sea contribution in the
magnetization density might be an indication of the importance of
the pion clouds effects in this quantity \cite{TTM81},\cite{HDM04}.
To pursuit the coordinate
space interpretation of the different $Q^2$-dependence of the
electric and magnetic form factors, it is preferable to compare the charge
density $\rho_{ch} (r)$ with the magnetization density $\rho_m^K (r)$
introduced by Kelly rather than with the magnetization density
$\rho_m (r)$ appearing in Eq.(\ref{eq:mgdens}).
The density $\rho_m^K (r)$ can easily be obtained from the theoretical
$\rho_m (r)$ through the relation (\ref{eq:mgrelation}).
We emphasize again that, since Eq.(\ref{eq:mgrelation}) directly relates
the two magnetization densities in the intrinsic frame or the nucleon
rest frame, the density $\rho_m^K (r)$ obtained in that way is
{\it completely free from the boost procedure}, especially from the
choice of the boost mass $M_B$ as well as the parameters
$\lambda_E$ and $\lambda_M$.

\begin{figure}[htb]
\begin{center}
  \includegraphics[height=.36\textheight]{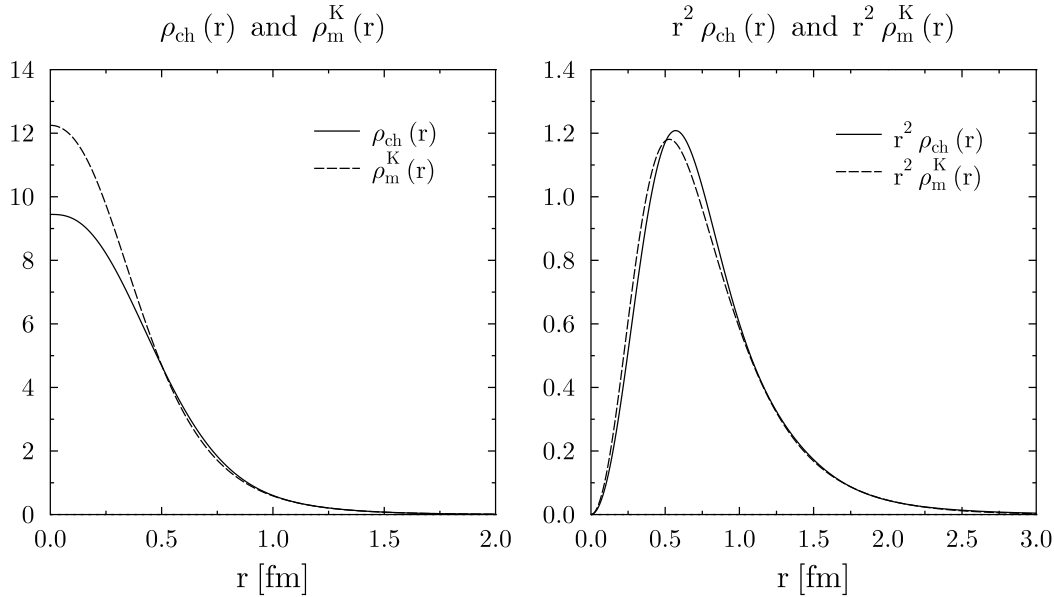}
  \caption{\baselineskip16pt The CQSM predictions for the charge,
  $\rho_{ch} (r)$, and Kelly's magnetization, $\rho_m^K (r)$, densities
  (left panel) and the corresponding $r^2$-weighted densities (right
  panel).}%
\label{Fig2}
\end{center}
\end{figure}

The left panel of Fig.\ref{Fig2} represents the theoretical charge and
Kelly's magnetization densities. Remember that the standard magnetization
density $\rho_m (r)$ shown in Fig.\ref{Fig1}
has an entirely different radial shape from the charge density
$\rho_{ch} (r)$. Nonetheless, the radial dependence of Kelly's magnetization
density $\rho_m^K (r)$, which is obtained from $\rho_m (r)$ through the
relation (\ref{eq:mgrelation}), is unexpectedly close to that
of $\rho_{ch} (r)$.
Still, we observe a noticeable qualitative difference between
$\rho_{ch} (r)$ and $\rho_m^K (r)$.
The proton charge density is a little broader than its
magnetization density, although the size of the difference predicted
by the CQSM might not be large enough as compared with the one
suggested by Kelly's phenomenological analysis. This tendency can
more clearly be seen by comparing the $r^2$-weighted
densities, i.e. $r^2 \,\rho_{ch} (r)$ and $r^2 \,\rho_m^K (r)$,
illustrated in the right panel of Fig.\ref{Fig2}.
One confirms that the
peak position of $r^2 \,\rho_m^K (r)$ is located at smaller $r$ than
the charge density $r^2 \,\rho_{ch} (r)$, in consistent with Kelly's
result shown in Fig.7 of \cite{Kelly02}.
Putting it another way, $r^2 \,\rho_m^K (r)$
is larger than $r^2 \,\rho_{ch} (r)$ in the inner region
$r \lesssim 0.5 \,\mbox{fm}$, while the converse is true in the range in
the range $0.5 \,\mbox{fm} \lesssim r \lesssim 1.0 \,\mbox{fm}$.
Undoubtedly, this qualitative difference between the intrinsic charge and
magnetization density must be the source of the observed fast decrease of
the ratio $G_E (Q^2) \,/ \,G_M (Q^2)$ at high-momentum transfer
in the JLab measurements.

To convince that the predicted delicate difference between the
intrinsic charge and magnetization densities is in fact an
essential ingredient to generate the observed difference
between the Sachs electric and magnetic form factors,
we next try to evaluate the latters by using a simple
prescription given by (1),(2),(3) and (4),(5).
For the parameters $\lambda_E$ and $\lambda_M$, we use 
here the simplest choice $\lambda_E = \lambda_M = 0$, which
amounts to taking in only the Lorentz contraction of the
spatial distributions of the constituents.
Then, only one remaining parameter of our analysis below is
the boost mass $M_B$, which ideally should coincide with the
physical nucleon mass, or the classical soliton mass
in our theoretical treatment.
Following the previous studies \cite{Holz96}\nocite{Hplz02}-\cite{Holz05},
here we treat it as additional parameter of the analysis.
It was determined to be $M_B \simeq 1.2 \,\mbox{GeV}$ so that it
reproduces the general trend of the $Q^2$-dependence of the ratio
$R \equiv \mu_p \,G_E (Q^2) \,/ \,G_M (Q^2)$.

\begin{figure}[htb]
\begin{center}
  \includegraphics[height=.38\textheight]{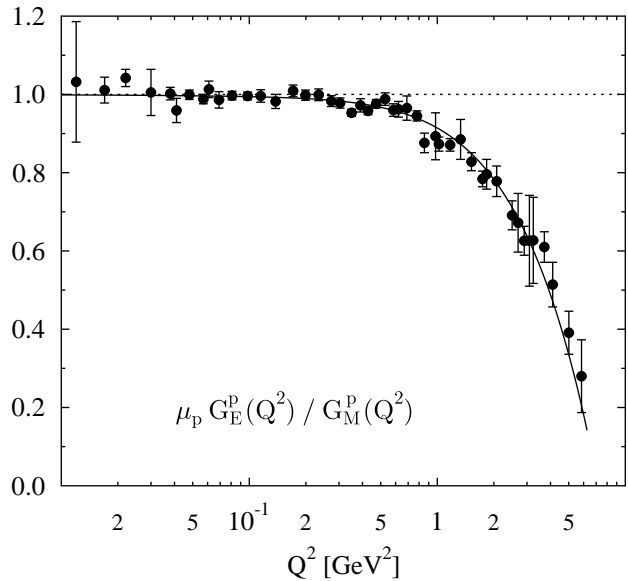}
  \caption{\baselineskip16pt The CQSM predictions for the form factor
  ratio $R \equiv \mu_p \,G_E^p (Q^2) \,/ \,G_M^p (Q^2)$}%
\label{Fig3}
\end{center}
\end{figure}

Fig.\ref{Fig3} shows the CQSM prediction for the ratio
$\mu_p \,G_E (Q^2) \,/ \,G_M (Q^2)$ obtained in the above way, in
comparison with the recent global fit obtained by taking account of
two-photon exchange contributions and their associated
uncertainties \cite{AMT07}.
As one can see, the agreement between the theoretical prediction and
the empirical data is pretty good.
(This agreement should not be overestimated too much,
however, since the gross $Q^2$-dependence beyond the range
$\sqrt{Q^2} \gtrsim M_N$ is largely controlled by our approximate
treatment of the Lorentz boost effects.)
A natural question here is why a fairly small difference between the
two intrinsic electromagnetic densities causes such a
significant difference between the $Q^2$-dependencies of the electric
and magnetic form factors in a few GeV region. As already indicated
in Holzwarth's analysis based on the generalized Skyrme
model \cite{Holz02},\cite{Holz05}, the existence of a zero in the
electric form factor may provide the simplest answer.

\begin{figure}[htb]
\begin{center}
  \includegraphics[height=.38\textheight]{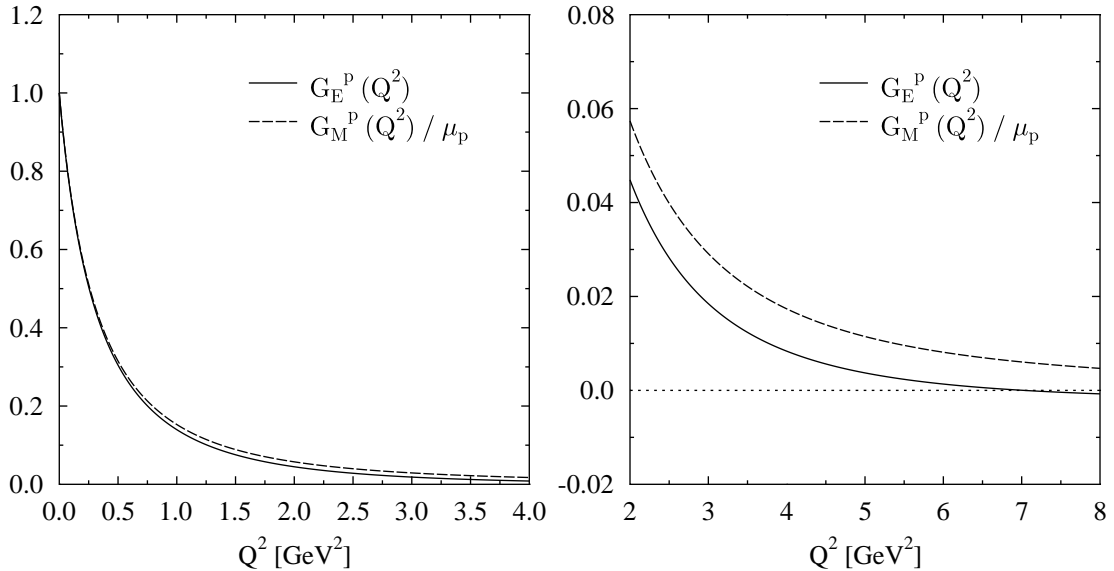}
  \caption{\baselineskip16pt Comparison of the CQSM predictions for 
  $G_E^p (Q^2)$ and $G_M^p (Q^2) \,/ \,\mu_p$. The right panel
  is a close-up picture of the high-momentum-transfer region.}%
\label{Fig4}
\end{center}
\end{figure}

To confirm it, we show in Fig.\ref{Fig4} the prediction of the present
model for $G_E (Q^2)$ and $G_M (Q^2)$. One sees that the first
zero of $G_E (Q^2)$ appears around $Q^2 \simeq 7 \,\mbox{GeV}^2$,
while $G_M (Q^2)$ still remains positive in the investigated
$Q^2$ range. Since the magnitude of the electric form factor
decreases rapidly as $Q^2$ approaches its zero, the fast falloff
of the ratio $R (Q^2) \equiv \mu_p G_E (Q^2) / G_M (Q^2)$
when coming close to this $Q^2$ region is easily understood.
Just to be sure, we never claim that our model is quantitative
enough to be able to predict precise position of the zero of the
electric form factor. Still, its qualitative prediction that the first
zero of the charge form factor appears at much lower momentum
transfer than that of the magnetic form factor would be intact and
that the cause of this feature can be traced back to the predicted
qualitative difference of the two intrinsic electromagnetic densities.
Now, admitting that the above interpretation on the strong
$Q^2$-dependence of the form factor ratio $R$ is correct, the
breakdown of the pQCD counting rule applied to this ratio seems only
natural. After all, zeros of the elastic form factors are outside
the applicability range of the pQCD counting rule.

\begin{figure}[htb]
\begin{center}
  \includegraphics[height=.38\textheight]{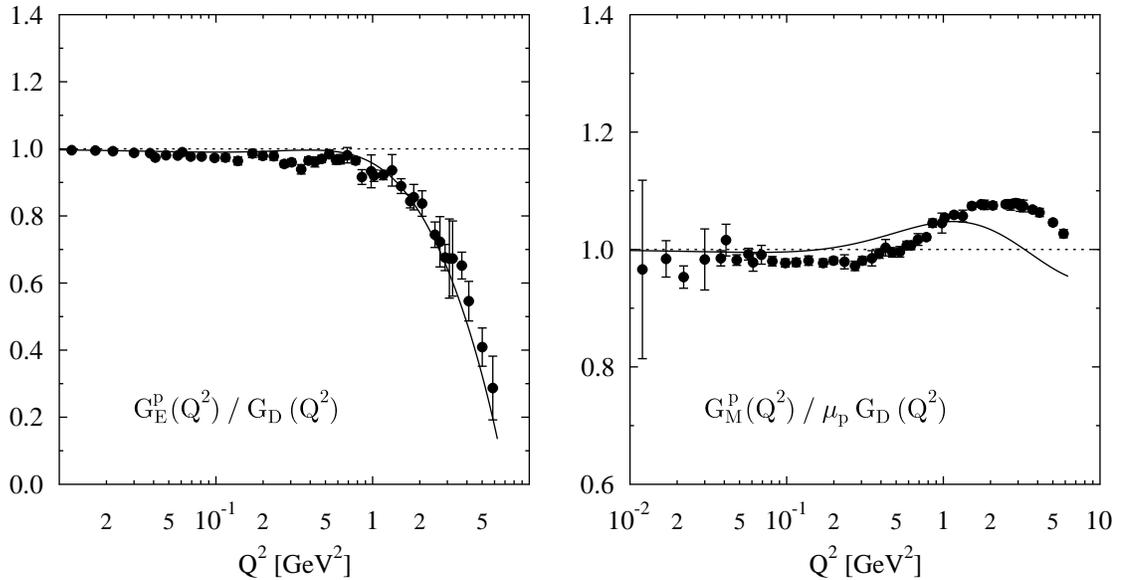}
  \caption{\baselineskip16pt The CQSM predictions for the ratios,
  $G_E^p (Q^2) \,/ \,G_D (Q^2)$ (left panel), and
  $G_M^p (Q^2) \,/ \,\mu_p \,G_D (Q^2)$ (right panel), where $G_D (Q^2)$
  is the dipole form factor with the dipole mass
  $M_D^2 = 0.62 \,\mbox{GeV}^2$. The empirical data are
  from \cite{AMT07}.}%
\label{Fig5}
\end{center}
\end{figure}

To show a subtle difference between the two form factors
$G_E (Q^2)$ and $G_M (Q^2)$, it is customary to compare
the ratio of each form factor to the dipole form factor
$G_D (Q^2) = \left( 1 + Q^2 / M_D^2 \right)^{-2}$.
The dipole mass $M_D$, which reproduce the gross $Q^2$-dependence
of the observed electric and magnetic form factors is known to be
$M_D^2 = 0.71 \,\mbox{GeV}^2$.
Unfortunately, the CQSM slightly overestimates the nucleon
electromagnetic sizes. Accordingly, the dipole mass, which reproduces
the average $Q^2$-dependence of the two theoretical form factors
$G_E (Q^2)$ and $G_M (Q^2)$ turns out to be a little smaller than the
standard one, i.e.
$M_D^2 \simeq 0.62 \,\mbox{GeV}^2$. The left and the right panels
of Fig.\ref{Fig5} respectively stand for the form factor ratios
$G_E (Q^2) / G_D (Q^2)$ and $G_M (Q^2) / G_D (Q^2)$ obtained with
this dipole mass.
Although the agreement between the theoretical predictions and the
empirical data is far from perfect, a remarkable qualitative difference
between the $Q^2$-dependence of
$G_E (Q^2) / G_D (Q^2)$ and $G_M (Q^2) / G_D (Q^2)$ is clearly
seen. Since we are using the same $\lambda$ parameters, i.e.
$\lambda_E = \lambda_M = 0$, and a common boost mass $M_B$ for obtaining
the electric and magnetic form factors, the cause of this difference
must purely be attributed to the difference of the intrinsic charge and
magnetization densities predicted by the CQSM.

To summarize, in pursuit of the physical origin of the observed behavior
of the form factor ratio $\mu_p \,G_E (Q^2) / G_M (Q^2)$ of the proton,
we have carried out a comparative analysis of the intrinsic charge and
magnetization densities defined in the nucleon rest frame, within the
framework of the CQSM containing only one parameter, i.e. the
dynamical quark mass.
The point is that our predictions for these two intrinsic densities
are completely free from the difficult problem of relativistic
recoil effects. We find that the predicted intrinsic charge density
of the proton is a little broader than the intrinsic magnetization
density {\it a la} Kelly in qualitatively consistent with Kelly's
conclusion obtained from the empirical information on the Sachs
electric and magnetic form factors.
This then gives a strong theoretical support to Kelly's
coordinate space interpretation of the faster falloff of the electric
form factor than the magnetic one. Just to be sure,
we have further checked that the predicted delicate difference
between the theoretical intrinsic charge and magnetization densities,
supplemented with a simple introduction of
Lorentz contraction effects, can reproduce the observed difference
between the charge and magnetic form factors up to
$Q^2 \simeq 6 \,\mbox{GeV}^2$.
It was indicated there that the existence of a zero of the
electric form factor $G_E (Q^2)$ around $Q^2 \simeq 7 \,\mbox{GeV}^2$
is a likely reason of remarkably fast decrease of the form
factor ratio $R \equiv \mu_p \,G_E (Q^2) / G_M (Q^2)$ beyond
$Q^2 \simeq 2 \,\mbox{GeV}^2$.
Although this latter part of study is of totally approximate nature and
should be taken as qualitative, we believe that our present analysis
all together has succeeded to give a valuable insight into the physics
behind the form factor ratio $\mu_p \,G_E^p (Q^2) / G_M^p (Q^2)$.

\vspace{10mm}
\noindent
{\bf Acknowledgement}

\vspace{2mm}
This work is supported in part by a Grant-in-Aid for Scientific
Research for Ministry of Education, Culture, Sports, Science
and Technology, Japan (No.~C-16540253)

\end{document}